\documentclass[draftclsnofoot,peerreview,english]{IEEEtran}
\usepackage[T1]{fontenc}
\usepackage[latin9]{inputenc}
\usepackage{amsmath}
\usepackage{amssymb}
\usepackage{mathrsfs}
\usepackage{bbm}
\usepackage{mathtools}
\usepackage{empheq}
\usepackage{graphicx}
\usepackage{amsthm}	
\usepackage{hyperref}
\usepackage{cite}
\usepackage{accents}


\usepackage{babel}

\newtheorem{thm}{Theorem}				
\newtheorem{prop}[thm]{Proposition}		
\newtheorem{lem}[thm]{Lemma}			
\newtheorem{cor}[thm]{Corollary}

\begin{document}

\title{Can Negligible Cooperation Increase Capacity?\\
The Average-Error Case}

\author{Parham~Noorzad, Michelle Effros, Michael
Langberg%
\thanks{This material is based upon work supported by the 
National Science Foundation under Grant Numbers 1527524 and
1526771.}%
\thanks{P. Noorzad was with the California
Institute of Technology, Pasadena, CA 91125 USA.
He is now with Qualcomm Research, San Diego, CA 
92122 USA (email: parham@qti.qualcomm.com). }%
\thanks{M. Effros is with the California
Institute of Technology, Pasadena, CA 91125 USA
(email: effros@caltech.edu). }%
\thanks{M. Langberg is with the State
University of New York at Buffalo, Buffalo, NY 14260 USA
(email: mikel@buffalo.edu).}}

\maketitle

\begin{abstract}
In communication networks, cooperative strategies
are coding schemes where network nodes work together 
to improve network performance metrics such as sum-rate.
This work studies \emph{encoder} cooperation
in the setting of a discrete multiple access channel with
two encoders and a single decoder. A node in the network
that is connected to both encoders via rate-limited links,
referred to as the cooperation facilitator (CF), enables 
the cooperation strategy. Previously, the authors 
presented a class of multiple access channels where the 
\emph{average-error} sum-capacity has an infinite derivative 
in the limit where CF output link capacities approach zero.
The authors also demonstrated that for some channels, 
the \emph{maximal-error} sum-capacity is not continuous 
at the point where the output link capacities of the CF equal zero.
This work shows that the the average-error sum-capacity
is continuous when CF output link capacities converge to zero; 
that is, the infinite derivative of the average-error
sum-capacity is not a result of its discontinuity as in the
maximal-error case. 
\end{abstract}


\section{Introduction} \label{sec:intro}

Interference is an important limiting factor
in the capacity performance of communication 
networks. One way to reduce interference is for 
nodes in the network to work together to coordinate
their transmissions. Any such strategy falls under
the general definition of cooperation.

\begin{figure} 
	\begin{center}
		\includegraphics[scale=0.25]{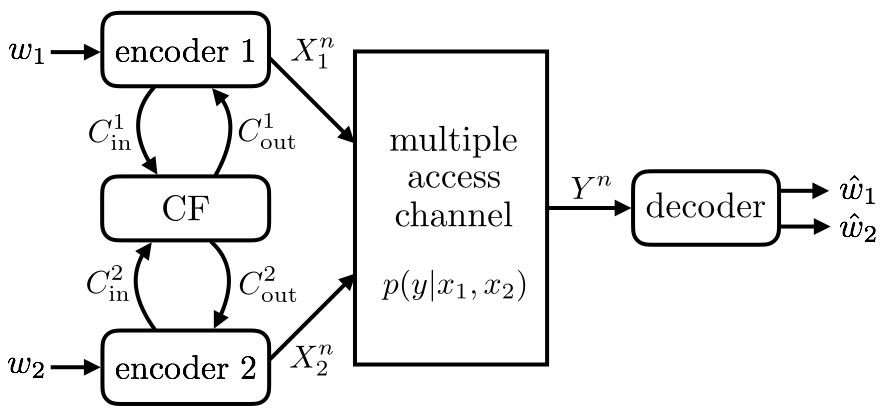}
		\caption{A network consisting of two encoders that 
		initially cooperate via a CF and then, based on the 
		information they receive, transmit their codewords
		over the MAC to the decoder.} \label{fig:model}
	\end{center}
\end{figure}

While strategies such as time-sharing are one form
of cooperation, the model we consider
here is closer to the ``conferencing''
cooperation model \cite{WillemsMAC}.
In conferencing, unlike time-sharing, 
encoders share information about the messages they 
wish to transmit. In contrast to conferencing however, 
our cooperation model employs \emph{indirect} communication;
that is, the encoders communicate through another
node which we call the cooperation facilitator
(CF) \cite{reliability, kUserMAC}. Figure \ref{fig:model} 
depicts the CF model in the
two-user multiple access channel (MAC) scenario.

The CF enables cooperation between the encoders through
its rate-limited input and output links. Prior to the transmission
of its codeword over the channel, each encoder sends a
function of its message to the CF. The CF, using
the information it receives from \emph{both} encoders, computes
a rate-limited function for each encoder. It then transmits the 
computed values over its output links. Each encoder then selects a
codeword using its message and the information it receives from
the CF.

To simplify our discussion in this section, suppose the CF input link capacities
both equal $C_\mathrm{in}$ and the CF output link capacities both
equal $C_\mathrm{out}$. If $C_\mathrm{in}\leq C_\mathrm{out}$, then
the optimal strategy for the CF is ``forwarding'' \cite{kUserMAC}; 
that is, the CF simply
forwards the information it receives from one encoder to the other.
Using the capacity region of the MAC with
conferencing encoders \cite{WillemsMAC}, it follows that the 
average-error sum-capacity gain of CF cooperation is 
bounded from above by $2C_\mathrm{in}$ and does not explicitly depend on 
$C_\mathrm{out}$. In the
situation where $C_\mathrm{in}>C_\mathrm{out}$, however, the situation 
is more complicated since the CF can no longer forward all its
incoming information. While the $2C_\mathrm{in}$ upper bound is still
valid, the dependence of the sum-capacity gain on $C_\mathrm{out}$ is
less clear. If the CF simply forwards part of the information it 
receives, then again by \cite{WillemsMAC}, the average-error sum-capacity
gain is at most $2C_\mathrm{out}$. While the $2C_\mathrm{out}$ bound has
an intuitive interpretation, in the sense that it reflects the amount
of information the CF shares with the encoders, a much larger gain 
is possible. Specifically, in prior work \cite{kUserMAC}, we
show that for a class of MACs which includes the binary adder
MAC,\footnote{The binary adder MAC defined as $Y=X_1+X_2$, 
where $X_1,X_2\in\{0,1\}$, and $Y\in\{0,1,2\}$.} for fixed $C_\mathrm{in}>0$,
the average-error sum-capacity has a derivative 
in $C_\mathrm{out}$ that is infinite at $C_\mathrm{out}=0$; that is, for small
$C_\mathrm{out}$, the gain resulting from cooperation exceeds any function
with bounded derivative. 

The large sum-capacity gain described above is not limited to 
the average-error scenario. In fact, in related work \cite{reliability}, 
we show that for any MAC for which the average-error sum-capacity
is strictly greater than the maximal-error sum-capacity in the absence of cooperation, 
adding a CF and measuring  
the \emph{maximal-error} sum-capacity for fixed $C_\mathrm{in}>0$ 
gives a curve that is discontinuous at $C_\mathrm{out}=0$.
In this case, we say that ``negligible cooperation'' results in a 
non-negligible benefit.\footnote{In \cite{SingleBit}, Langberg
and Effros further show that in a certain network, the transmission of 
even a single bit (over the entire blocklength) can 
strictly increase maximal-error capacity.}

Given these earlier results, the following question remains open: For fixed
$C_\mathrm{in}>0$, can the \emph{average-error} sum-capacity
ever be discontinuous in $C_\mathrm{out}$? In the present work, 
we show that the answer is negative; that is, the average-error 
sum-capacity is continuous even at $C_\mathrm{out}=0$. 

A related average-error continuity problem is the 
case of rate-limited feedback over the MAC. For such a network, 
Sarwate and Gastpar \cite{SarwateGastpar}, using the 
dependence-balance bounds of Hekstra and Willems
\cite{HekstraWillems}, show that as the feedback rate 
converges to zero, the average-error capacity region 
converges to the average-error capacity region of 
the same MAC in the absence of feedback. 

The problem we study here can also be formulated as an
``edge removal problem'' as introduced by Ho, Effros, and
Jalali \cite{HoEtAl, JalaliEtAl}. 
The edge removal problem seeks to quantify the capacity 
effect of removing a single edge from a network. 
Except in the MAC with CF setting and a number of cases 
described in \cite{HoEtAl, JalaliEtAl}, the problem remains open. 
In fact, Effros and Langberg show that this problem is 
connected to a number of other open problems in 
network coding, including the difference between the 
$0$-error and $\epsilon$-error capacity regions \cite{LangbergEffros1}
and the difference between the lossless source coding regions for
independent and dependent sources \cite{LangbergEffros2}. 

The question of whether the capacity region of a network consisting
of noiseless links is continuous with respect to the link capacities
is investigated by Gu, Effros, and Bakshi \cite{GuEffrosBakshi}
and Chan and Grant \cite{ChanGrant}. The present work differs from
\cite{GuEffrosBakshi, ChanGrant} in the network under consideration;
while our network does have noiseless links (the CF input and output 
links), it also contains a multiterminal component (the MAC) which 
may exhibit interference or noise. 

In \cite{KosutKliewer}, Kosut and Kliewer present different variations
of the edge removal problem in a unified setting. In their terminology,
the present work investigates whether the network consisting of a MAC 
and a CF satisfies the ``weak edge removal property'' with respect 
to the average-error reliability criterion. In \cite[Chapter 1]{NoorzadPhD},
we discuss the known results for each variation of the edge removal 
problem. 

\section{The Cooperation Facilitator Model} \label{sec:model}

In this work, we study cooperation between two encoders that 
communicate their messages to a decoder 
over a discrete, memoryless, and stationary MAC.
Such a MAC can be represented by the triple
\begin{equation*}
  \big(\mathcal{X}_1\times\mathcal{X}_2,p(y|x_1,x_2),
  \mathcal{Y}\big),
\end{equation*}
where $\mathcal{X}_1$, $\mathcal{X}_2$, and $\mathcal{Y}$ 
are finite sets and $p(y|x_1,x_2)$ is a 
conditional probability mass function. 
The $n$th extension of the channel is given by
\begin{equation*}
  p(y^n|x_1^n,x_2^n)\coloneqq
  \prod_{t=1}^n p(y_t|x_{1t},x_{2t}).
\end{equation*}

For each positive integer $n$, called the
``blocklength,'' and nonnegative real numbers $R_1$ and
$R_2$, called the ``rates,'' we next define a 
\begin{equation*}
  \big(2^{nR_1},2^{nR_2},n\big)\text{-code}
\end{equation*}
for communication over a MAC with a 
$(\mathbf{C}_\mathrm{in},\mathbf{C}_\mathrm{out})$-CF. Here 
$\mathbf{C}_\mathrm{in}=(C_\mathrm{in}^1,C_\mathrm{in}^2)$ and 
$\mathbf{C}_\mathrm{out}=(C_\mathrm{out}^1,C_\mathrm{out}^2)$
represent the capacities of the CF input and output links,
respectively. (See Figure \ref{fig:model}.)

For $i\in\{1,2\}$, the transmission of encoder $i$ to 
the CF is represented by a mapping\footnote{Henceforth, 
for every $x\geq 1$, $[x]$ denotes the set $\{1,\dots,\lfloor x\rfloor\}$.}
\begin{equation*} 
  \varphi_i\colon [2^{nR_i}]
  \rightarrow [2^{nC_\mathrm{in}^i}].
\end{equation*}
The CF, based on the information it receives from the 
encoders, computes a function
\begin{equation*} 
  \psi_i\colon [2^{nC_\mathrm{in}^1}]\times [2^{nC_\mathrm{in}^2}]
  \rightarrow [2^{nC_\mathrm{out}^i}]
\end{equation*}
for encoder $i$, where $i\in\{1,2\}$.
Encoder $i$, using its message and what it
receives from the CF, selects a codeword according to
\begin{equation*}
  f_i\colon [2^{nR_i}]\times [2^{nC_\mathrm{out}^i}]
  \rightarrow \mathcal{X}_i^n.
\end{equation*}
The decoder, using the channel output, aims to find the transmitted
messages. It is represented by a mapping
\begin{equation*}
  g\colon\mathcal{Y}^n\rightarrow
  [2^{nR_1}]\times [2^{nR_2}].
\end{equation*}
The collection of mappings 
\begin{equation*}
  \big(\varphi_1,\varphi_2,\psi_1,\psi_2,
  f_1,f_2,g\big)
\end{equation*}
defines a $(2^{nR_1},2^{nR_2},n)$-code for the MAC
with a $(\mathbf{C}_\mathrm{in},\mathbf{C}_\mathrm{out})$-CF.\footnote{Technically,
the definition we present here is for a single round of cooperation.
Similar to \cite{reliability}, it is possible to define cooperation 
via a CF over multiple rounds. However, this general scenario
does not alter our main proofs.}

For a fixed code, the error probability for a particular
message pair $(w_1,w_2)$ is given by
\begin{equation*}
  \lambda_n(w_1,w_2)\coloneqq
  \sum_{y^n\colon g(y^n)\neq (w_1,w_2)}
  p(y^n|f_1(w_1,z_1),f_2(w_2,z_2)),
\end{equation*}
where $z_1$ and $z_2$ are the CF outputs. For 
$i\in\{1,2\}$, $z_i$ is calculated according to
\begin{equation*}
  z_i = \psi_i\big(\varphi_1(w_1),\varphi_2(w_2)\big).
\end{equation*}
The \emph{average} probability of error is given by
\begin{equation*}
  P_{e,\mathrm{avg}}^{(n)}\coloneqq
  \frac{1}{2^{n(R_1+R_2)}}\sum_{w_1,w_2}
  \lambda_n(w_1,w_2),
\end{equation*}
and the \emph{maximal} probability of error is given by
\begin{equation*}
  P_{e,\mathrm{max}}^{(n)}\coloneqq
  \max_{w_1,w_2} \lambda_n(w_1,w_2).
\end{equation*}

A rate pair $(R_1,R_2)$ is achievable with respect to 
the average-error reliability criterion if there exists a sequence of 
$(2^{nR_1},2^{nR_2},n)$-codes such that 
$P^{(n)}_{e,\mathrm{avg}}\rightarrow 0$ as 
$n\rightarrow\infty$. The average-error capacity region
of a MAC with a 
$(\mathbf{C}_\mathrm{in},\mathbf{C}_\mathrm{out})$-CF, denoted by 
$\mathscr{C}_\mathrm{avg}(\mathbf{C}_\mathrm{in},\mathbf{C}_\mathrm{out})$,
is the closure of the set of all rate pairs
that are achievable with respect to the average-error reliability
criterion. The average-error sum-capacity is defined as 
\begin{equation*}
  C_\mathrm{sum,avg}(\mathbf{C}_\mathrm{in},\mathbf{C}_\mathrm{out})
  \coloneqq 
  \max_{\mathscr{C}_\mathrm{avg}(\mathbf{C}_\mathrm{in},\mathbf{C}_\mathrm{out})} 
  (R_1+R_2)
\end{equation*}

By replacing $P^{(n)}_{e,\mathrm{avg}}$ with 
$P^{(n)}_{e,\mathrm{max}}$, we can similarly define
achievable rates with respect to the maximal-error reliability criterion, 
the maximal-error capacity region, and the maximal-error 
sum-capacity. For a MAC with a 
$(\mathbf{C}_\mathrm{in},\mathbf{C}_\mathrm{out})$-CF,
we denote the maximal-error capacity region 
and sum-capacity by 
$\mathscr{C}_\mathrm{max}(\mathbf{C}_\mathrm{in},\mathbf{C}_\mathrm{out})$
and
$C_\mathrm{max}(\mathbf{C}_\mathrm{in},\mathbf{C}_\mathrm{out})$,
respectively.

\section{Prior Results on the
Sum-Capacity Gain of Cooperation} \label{sec:knownApriori}

We next describe results from 
\cite{reliability, kUserMAC} that are relevant to our 
discussion here. We begin with sum-capacity results
in the average-error case.

Consider a discrete MAC 
$(\mathcal{X}_1\times\mathcal{X}_2,p(y|x_1,x_2),\mathcal{Y})$.
Let $p_\mathrm{ind}(x_1)p_\mathrm{ind}(x_2)$ be a distribution
that satisfies
\begin{equation} \label{eq:pind}
  I_\mathrm{ind}(X_1,X_2;Y)
  =\max_{p(x_1)p(x_2)}I(X_1,X_2;Y).
\end{equation}
In addition, suppose that there exists a distribution 
$p_\mathrm{dep}(x_1,x_2)$ whose support is contained in
the support of $p_\mathrm{ind}(x_1)p_\mathrm{ind}(x_2)$, 
and satisfies 
\begin{equation} \label{eq:pindApdep}
  I_\mathrm{dep}(X_1,X_2;Y)
  + D\big(p_\mathrm{dep}(y)\|p_\mathrm{ind}(y)\big)
  > I_\mathrm{ind}(X_1,X_2;Y).
\end{equation}
Let $\mathcal{C}^*$ denote the class of all discrete 
MACs for which input distributions $p_\mathrm{ind}$
and $p_\mathrm{dep}$, as described above, exist. Then
the following theorem \cite[Theorem 3]{kUserMAC} holds.
\begin{thm} \label{thm:infSlope}
Let 
$(\mathcal{X}_1\times\mathcal{X}_2,p(y|x_1,x_2),\mathcal{Y})$
be a MAC in $\mathcal{C}^*$, and suppose 
$(\mathbf{C}_\mathrm{in},\mathbf{v})
\in\mathbb{R}_{>0}^2\times \mathbb{R}_{>0}^2$.
Then
\begin{equation*}
  \lim_{h\rightarrow 0^+}
  \frac{C_\mathrm{sum,avg}(\mathbf{C}_\mathrm{in},h\mathbf{v})
  -C_\mathrm{sum,avg}(\mathbf{C}_\mathrm{in},\mathbf{0})}{h}
  =\infty.
\end{equation*}
\end{thm}

Consider a MAC in $\mathcal{C}^*$ with a CF that
has input links with equal capacity $C_\mathrm{in}$
and output links with equal capacity $C_\mathrm{out}$. Then
Theorem \ref{thm:infSlope} implies that for fixed $C_\mathrm{in}>0$,
the average-error sum-capacity has a derivative in 
$C_\mathrm{out}$ that is infinite at $C_\mathrm{out}=0$.\footnote{Note that
Theorem 1 does not lead to any conclusions regarding continuity;
a function $f(x)$ with infinite derivative at $x=0$ can be continuous 
(e.g., $f(x)=\sqrt{x}$) or discontinuous 
(e.g., $f(x)=\lceil x\rceil)$.}

We next describe the maximal-error sum-capacity gain.
While it is possible in the average-error scenario 
to achieve a sum-capacity that has an infinite slope, 
more is known in the maximal-error case. There exists a class
of MACs for which the maximal-error sum-capacity 
exhibits a discontinuity in the capacities of the
CF output links. This is stated formally in the next 
proposition, which is a special case of 
\cite[Proposition 5]{reliability}.

\begin{prop} \label{prop:discontinuity}
Consider a discrete MAC for which\footnote{Dueck \cite{Dueck}
gives the first proof of the existence of a discrete MAC which satisfies
(\ref{eq:avgGeqMax}). We investigate further properties of
Dueck's MAC in \cite{reliability}.}
\begin{equation} \label{eq:avgGeqMax}
  C_\mathrm{sum,avg}(\mathbf{0},\mathbf{0})
  > C_\mathrm{sum,max}(\mathbf{0},\mathbf{0});
\end{equation}
that is, the average-error sum-capacity is strictly
greater than the maximal-error sum-capacity. 
Fix $\mathbf{C}_\mathrm{in}\in\mathbb{R}^2_{>0}$.
Then 
$C_\mathrm{sum,max}(\mathbf{C}_\mathrm{in},
\mathbf{C}_\mathrm{out})$
is not continuous at 
$\mathbf{C}_\mathrm{out}=\mathbf{0}$. 
\end{prop}

The possibility of an infinite derivative in the average-error case
(Theorem \ref{thm:infSlope}) and a discontinuity in the maximal-error scenario
(Proposition \ref{prop:discontinuity}) leads to the following question:
Does there exist any MAC and any $\mathbf{C}_\mathrm{in}$
for which 
$C_\mathrm{sum,avg}(\mathbf{C}_\mathrm{in},\mathbf{C}_\mathrm{out})$ 
is not continuous at $\mathbf{C}_\mathrm{out}=\mathbf{0}$? This
problem is posed in \cite[Section IV]{reliability}. 
We address this question in the next section, where we
describe our results. 

\section{Continuity of Average- and Maximal-Error Sum-Capacities} 
\label{sec:results}

In the prior section, for a fixed $\mathbf{C}_\mathrm{in}$,
we discuss previous results regarding the continuity
of $C_\mathrm{sum,avg}(\mathbf{C}_\mathrm{in},\mathbf{C}_\mathrm{out})$
and $C_\mathrm{sum,max}(\mathbf{C}_\mathrm{in},\mathbf{C}_\mathrm{out})$
as a function of $\mathbf{C}_\mathrm{out}$ at
$\mathbf{C}_\mathrm{out}=\mathbf{0}$. In this section, we do not
limit ourselves to the point $\mathbf{C}_\mathrm{out}=\mathbf{0}$;
rather, we study the continuity of 
$C_\mathrm{sum,avg}(\mathbf{C}_\mathrm{in},\mathbf{C}_\mathrm{out})$
over its entire domain. 
 
We begin by considering the case where the CF has full
access to the messages. Formally, for a given discrete MAC 
$(\mathcal{X}_1\times\mathcal{X}_2,p(y|x_1,x_2),\mathcal{Y})$,
let the components of 
$\mathbf{C}_\mathrm{in}^*=(C_\mathrm{in}^{*1},C_\mathrm{in}^{*2})$ 
be sufficiently large so that 
any CF with input link capacities $C_\mathrm{in}^{*1}$ and
$C_\mathrm{in}^{*2}$ has 
full knowledge of the encoders' messages. For example,
we can choose $\mathbf{C}_\mathrm{in}^*$ such that
\begin{equation*}
  \min\{C_\mathrm{in}^{*1},C_\mathrm{in}^{*2}\}
  >\max_{p(x_1,x_2)}I(X_1,X_2;Y).
\end{equation*}
Our first result addresses the continuity of 
$C_\mathrm{sum,avg}(\mathbf{C}_\mathrm{in}^*,\mathbf{C}_\mathrm{out})$
as a function of $\mathbf{C}_\mathrm{out}$ over 
$\mathbb{R}^2_{\geq 0}$.

\begin{thm} \label{thm:CsumAvgContinuity}
For any discrete MAC, the mapping
\begin{equation} \label{eq:CsumAvgMap}
  \mathbf{C}_\mathrm{out}
  \mapsto C_\mathrm{sum,avg}(\mathbf{C}_\mathrm{in}^*,
  \mathbf{C}_\mathrm{out}),
\end{equation}
defined on $\mathbb{R}^2_{\geq 0}$ is continuous.
\end{thm}

We provide an overview of the proof in 
Section \ref{sec:fullKnowledgeCF} and present the
details in Section \ref{sec:proofs}. 

While Theorem \ref{thm:CsumAvgContinuity} focuses on the scenario
where $\mathbf{C}_\mathrm{in}=\mathbf{C}_\mathrm{in}^*$,
its result is sufficiently strong to address the continuity
problem for a fixed, arbitrary $\mathbf{C}_\mathrm{in}$
at $\mathbf{C}_\mathrm{out}=\mathbf{0}$. To see this,
note that for all $\mathbf{C}_\mathrm{in}\in\mathbb{R}^2_{\geq 0}$, 
\begin{equation} \label{eq:corAtZeroProof}
  C_\mathrm{sum,avg}(\mathbf{C}_\mathrm{in},
  \mathbf{0}) \leq
  C_\mathrm{sum,avg}(\mathbf{C}_\mathrm{in},
  \mathbf{C}_\mathrm{out})
  \leq 
  C_\mathrm{sum,avg}(\mathbf{C}_\mathrm{in}^*,
  \mathbf{C}_\mathrm{out}).
\end{equation}
Corollary \ref{cor:continuityAtZero}, below, now
follows from Theorem \ref{thm:CsumAvgContinuity}
by letting $\mathbf{C}_\mathrm{out}$ approach zero
in (\ref{eq:corAtZeroProof}) and noting that for all
$\mathbf{C}_\mathrm{in}\in\mathbb{R}^2_{\geq 0}$,
\begin{equation*}
  C_\mathrm{sum,avg}(\mathbf{C}_\mathrm{in}^*,\mathbf{0})
  =C_\mathrm{sum,avg}(\mathbf{C}_\mathrm{in},\mathbf{0})
  =C_\mathrm{sum,avg}(\mathbf{0},\mathbf{0}).
\end{equation*}

\begin{cor} \label{cor:continuityAtZero}
For any discrete MAC and any 
$\mathbf{C}_\mathrm{in}\in\mathbb{R}^2_{\geq 0}$, 
the mapping
\begin{equation*} 
  \mathbf{C}_\mathrm{out}
  \mapsto C_\mathrm{sum,avg}(\mathbf{C}_\mathrm{in},
  \mathbf{C}_\mathrm{out}),
\end{equation*}
is continuous at $\mathbf{C}_\mathrm{out}=\mathbf{0}$.
\end{cor}

Recall that Proposition \ref{prop:discontinuity} gives
a sufficient condition under which
$C_\mathrm{sum,max}(\mathbf{C}_\mathrm{in},\mathbf{C}_\mathrm{out})$
is not continuous at $\mathbf{C}_\mathrm{out}=\mathbf{0}$
for a fixed 
$\mathbf{C}_\mathrm{in}\in\mathbb{R}^2_{>0}$.
From Corollary \ref{cor:continuityAtZero}, it follows that
the sufficient condition is necessary as well. This is 
stated in the next corollary. The proof appears in 
Section \ref{sec:proofs}. 

\begin{cor} \label{cor:discontinuity}
Fix a discrete MAC and
$\mathbf{C}_\mathrm{in}\in\mathbb{R}^2_{>0}$.
Then 
$C_\mathrm{sum,max}(\mathbf{C}_\mathrm{in},
\mathbf{C}_\mathrm{out})$
is not continuous at 
$\mathbf{C}_\mathrm{out}=\mathbf{0}$ if and only if
\begin{equation}  \label{eq:avgBiggerThanMax}
  C_\mathrm{sum,avg}(\mathbf{0},\mathbf{0})
  > C_\mathrm{sum,max}(\mathbf{0},\mathbf{0}).
\end{equation}
\end{cor}

We next describe the second main result of this paper.
Our first main result, Theorem \ref{thm:CsumAvgContinuity},
shows the continuity of $C_\mathrm{sum,avg}(\mathbf{C}_\mathrm{in}^*,
\mathbf{C}_\mathrm{out})$ over $\mathbb{R}^2_{\geq 0}$.
The next result shows that proving the continuity
of $C_\mathrm{sum,avg}(\mathbf{C}_\mathrm{in},
\mathbf{C}_\mathrm{out})$ over 
$\mathbb{R}^2_{\geq 0}\times \mathbb{R}^2_{\geq 0}$
is equivalent to demonstrating its continuity on certain
axes. Specifically, it suffices to check the continuity
of $C_\mathrm{sum,avg}$ when one of $C_\mathrm{out}^1$
and $C_\mathrm{out}^2$ is approaching zero, while the
other arguments of $C_\mathrm{sum,avg}$ are fixed
positive numbers. 

\begin{thm} \label{thm:limitedCinContinuity}
For any discrete MAC, the mapping
\begin{equation} \label{eq:CsumAvgMap}
  (\mathbf{C}_\mathrm{in},\mathbf{C}_\mathrm{out})
  \mapsto C_\mathrm{sum,avg}(\mathbf{C}_\mathrm{in},
  \mathbf{C}_\mathrm{out}),
\end{equation}
defined on $\mathbb{R}^2_{\geq 0}\times\mathbb{R}^2_{\geq 0}$ is continuous
if and only if for all 
$(\mathbf{C}_\mathrm{in},\mathbf{C}_\mathrm{out})
\in\mathbb{R}^2_{>0}\times\mathbb{R}^2_{>0}$, we have
\begin{align*}
  \lim_{\tilde{C}_\mathrm{out}^1\rightarrow 0^+}
  C_\mathrm{sum,avg}\big(\mathbf{C}_\mathrm{in},
  (\tilde{C}_\mathrm{out}^1,C_\mathrm{out}^2)\big)
  &=C_\mathrm{sum,avg}\big(\mathbf{C}_\mathrm{in},
  (0,C_\mathrm{out}^2)\big)\\
  \lim_{\tilde{C}_\mathrm{out}^2\rightarrow 0^+}
  C_\mathrm{sum,avg}\big(\mathbf{C}_\mathrm{in},
  (C_\mathrm{out}^1,\tilde{C}_\mathrm{out}^2)\big)
  &=C_\mathrm{sum,avg}\big(\mathbf{C}_\mathrm{in},
  (C_\mathrm{out}^1,0)\big).
\end{align*}
\end{thm}

We remark that using a time-sharing code, 
it is possible to show that 
$C_\mathrm{sum,avg}$ is concave on 
$\mathbb{R}^2_{\geq 0}\times\mathbb{R}^2_{\geq 0}$, and thus
continuous on its interior. Thus it suffices to study the 
continuity of $C_\mathrm{sum,avg}$ on the boundary of 
$\mathbb{R}^2_{\geq 0}\times\mathbb{R}^2_{\geq 0}$. 

Sections \ref{sec:fullKnowledgeCF} and \ref{sec:arbitraryCF}
provide proof overviews of Theorem \ref{thm:CsumAvgContinuity} 
and Theorem \ref{thm:limitedCinContinuity},
respectively. Detailed proofs of the lemmas presented in 
these sections appear in Section \ref{sec:proofs}.

\section{Continuity of Sum-Capacity: The 
$\mathbf{C}_\mathrm{in}=\mathbf{C}_\mathrm{in}^*$
Case} \label{sec:fullKnowledgeCF}

We start our study of the continuity of $C_\mathrm{sum}(\mathbf{C}_\mathrm{in}^*,
\mathbf{C}_\mathrm{out})$\footnote{Henceforth, we write the average-error sum-capacity as
$C_\mathrm{sum}(\mathbf{C}_\mathrm{in}^*,
\mathbf{C}_\mathrm{out})$, since we are no longer concerned 
with the maximal-error sum-capacity.} by presenting lower and upper 
bounds in terms of an auxiliary function $\sigma(\delta)$ 
defined for $\delta\geq 0$
(Lemma \ref{lem:CsumAvgBounds}). 
A similar function appears in Dueck \cite{DueckSC}; our function 
differs with \cite{DueckSC} in a time-sharing random variable.
This random variable, denoted by $U$ below, plays two roles. First it ensures
$\sigma$ is concave, which immediately proves the continuity of
$\sigma$ over $\mathbb{R}_{>0}$. Second, together with a 
lemma from \cite{DueckSC} (Lemma \ref{lem:Dueck} below), it helps us 
find a single-letter upper bound
for $\sigma$ (Corollary \ref{cor:Dueck}). We then use the 
single-letter upper bound to prove continuity at $\delta=0$. 

The following definitions are useful for the description of 
our lower and upper bounds for 
$C_\mathrm{sum}(\mathbf{C}_\mathrm{in}^*,
\mathbf{C}_\mathrm{out})$.
For every finite alphabet $\mathcal{U}$ and all $\delta\geq 0$, define
\begin{equation*}
  \mathcal{P}^{(n)}_\mathcal{U}(\delta)
  :=\Big\{p(u,x_1^n,x_2^n)\Big |I(X_1^n;X_2^n|U)\leq n\delta\Big\}.
\end{equation*}
For every $n$, define the function 
$\sigma_n\colon\mathbb{R}_{\geq 0}\rightarrow\mathbb{R}_{\geq 0}$
as\footnote{For $n=1$, this function also appears in 
the study of the MAC with negligible feedback \cite{SarwateGastpar}.} 
\begin{equation} \label{eq:fndelta}
  \sigma_n(\delta)\coloneqq\sup_{\mathcal{U}}\max_{p\in \mathcal{P}^{(n)}_\mathcal{U}(\delta)}
  \frac{1}{n}I(X_1^n,X_2^n;Y^n|U),
\end{equation}
where the supremum is over all finite sets $\mathcal{U}$. As we see in
Lemma \ref{lem:concavity}, the conditioning on the random variable
$U$ in (\ref{eq:fndelta}) ensures that $\sigma_n$ is convex. 

For every $\delta\geq 0$, $(\sigma_n(\delta))_{n=1}^\infty$ satisfies 
a superadditivity property which appears in Lemma \ref{lem:superadditivity},
below. Intuitively, this property says that the sum-rate
of the best code of blocklength $m+n$ is bounded from below 
by the sum-rate of the concatenation of the best codes of
blocklengths $m$ and $n$.

\begin{lem} \label{lem:superadditivity}
For all $m,n\geq 1$ and all $\delta\geq 0$,
\begin{equation*}
  (m+n)\sigma_{m+n}(\delta)\geq m\sigma_m(\delta)
  +n\sigma_n(\delta).
\end{equation*}
\end{lem}

Given Lemma \ref{lem:superadditivity}, 
\cite[Appendix 4A, Lemma 2]{Gallager} now implies that
the sequence 
$(\sigma_n(\delta))_{n=1}^\infty$ 
converges for every $\delta\geq 0$,
and 
\begin{equation*}
  \lim_{n\rightarrow\infty}\sigma_n(\delta)
  =\sup_n \sigma_n(\delta).
\end{equation*}
Therefore, we can define the function 
$\sigma\colon\mathbb{R}_{\geq 0}\rightarrow\mathbb{R}_{\geq 0}$
as 
\begin{equation} \label{eq:FofDelta}
  \sigma(\delta)\coloneqq\lim_{n\rightarrow \infty}\sigma_n(\delta).
\end{equation} 
We next present our lower and upper bounds for 
$C_\mathrm{sum}(\mathbf{C}_\mathrm{in}^*,
\mathbf{C}_\mathrm{out})$ in terms of $\sigma$. The lower
bound follows directly from \cite[Corollary 8]{kUserMAC}.
\begin{lem} \label{lem:CsumAvgBounds}
For any discrete MAC and any $\mathbf{C}_\mathrm{out}\in\mathbb{R}^2_{\geq 0}$, 
we have
\begin{equation*}
  \sigma(C_\mathrm{out}^1+C_\mathrm{out}^2)
  -\min\{C_\mathrm{out}^1,C_\mathrm{out}^2\}\leq
  C_\mathrm{sum}(\mathbf{C}_\mathrm{in}^*,
  \mathbf{C}_\mathrm{out})\leq
  \sigma(C_\mathrm{out}^1+C_\mathrm{out}^2).
\end{equation*}
\end{lem}
From the remark following Theorem \ref{thm:limitedCinContinuity},
we only need to prove that 
$C_\mathrm{sum}(\mathbf{C}_\mathrm{in}^*,\mathbf{C}_\mathrm{out})$
is continuous on the boundary of $\mathbb{R}^2_{\geq 0}$. Note that
on the boundary of $\mathbb{R}^2_{\geq 0}$, however, 
$\min\{C_\mathrm{out}^1,C_\mathrm{out}^2\}=0$. Thus it suffices
to show that $\sigma$ is continuous on $\mathbb{R}_{\geq 0}$,
which is stated in the next lemma. 

\begin{lem} \label{lem:continuityOfF}
For any finite alphabet MAC, 
the function $\sigma$, defined by (\ref{eq:FofDelta}), 
is continuous on $\mathbb{R}_{\geq 0}$. 
\end{lem}

To prove Lemma \ref{lem:continuityOfF}, we first consider the continuity of $\sigma$
on $\mathbb{R}_{>0}$ and then deal with the point $\delta=0$. 
Note that $\sigma$ is the pointwise limit
of the sequence of functions $(\sigma_n)_{n=1}^\infty$.
Using a time-sharing argument as in 
\cite{CoverElGamalSalehi}, it is possible to 
show that each $\sigma_n$ is concave 
(Lemma \ref{lem:concavity}).
Therefore, $\sigma$ is concave as well, and
since $\mathbb{R}_{>0}$ is open, $\sigma$ is 
continuous on $\mathbb{R}_{>0}$.

\begin{lem}[Concavity of $\sigma_n$] 
\label{lem:concavity}
For all $n\geq 1$, $\sigma_n$ is concave 
on $\mathbb{R}_{\geq 0}$. 
\end{lem}

To prove the continuity of $\sigma(\delta)$
at $\delta=0$, we find an upper bound
for $\sigma$ in terms of $\sigma_1$. For
some finite set $\mathcal{U}$ and $\delta>0$,
consider a distribution 
$p\in\mathcal{P}^{(n)}_\mathcal{U}(\delta)$.
With respect to $p$, 
\begin{equation} \label{eq:multiletterMIbound}
  I(X_1^n;X_2^n|U)\leq n\delta.
\end{equation}
To find a bound for $\sigma$ in terms of $\sigma_1$,
we need a single-letter version of 
(\ref{eq:multiletterMIbound}). 
In \cite{DueckSC}, Dueck presents such 
a bound. We present Dueck's result in the
next lemma and include its proof for completeness. 

\begin{lem}[Dueck's Lemma \cite{DueckSC}] \label{lem:Dueck}
Fix $\epsilon,\delta >0$, positive integer $n$, and finite
alphabet $\mathcal{U}$. If $p\in\mathcal{P}^{(n)}_\mathcal{U}(\delta)$, then
there exists a set $T\subseteq [n]$ satisfying 
\begin{equation*}
  |T|\leq \frac{n\delta}{\epsilon},
\end{equation*}
such that 
\begin{equation*}
  \forall\: t\notin T\colon
  I(X_{1t};X_{2t}|U,X_1^T,X_2^T)\leq\epsilon,
\end{equation*}
where for $i\in\{1,2\}$, $X_i^T\coloneqq (X_{it})_{t\in T}$.
\end{lem}

Using Lemma \ref{lem:Dueck}, it is possible to find an upper
bound for $\sigma$ in terms of $\sigma_1$, which we present in 
the next corollary. The proof of this
corollary combines ideas from \cite{DueckSC} with results derived 
here. 

\begin{cor} \label{cor:Dueck}
For all $\epsilon,\delta >0$, we have
\begin{equation*}
  \sigma\big(\delta\big)\leq 
  \frac{\delta}{\epsilon}\log 
  |\mathcal{X}_1||\mathcal{X}_2|
  +\sigma_1(\epsilon).
\end{equation*}
\end{cor}

By Corollary \ref{cor:Dueck}, we have
\begin{equation*}
  \sigma(0)\leq \lim_{\delta\rightarrow 0^+}\sigma(\delta)
  \leq \sigma_1(\epsilon).
\end{equation*}
If we calculate the limit $\epsilon\rightarrow 0^+$, we get
\begin{equation*}
  \sigma(0)\leq \lim_{\delta\rightarrow 0^+}\sigma(\delta)
  \leq \lim_{\epsilon\rightarrow 0^+}\sigma_1(\epsilon).
\end{equation*}
Since $\sigma(0)=\sigma_1(0)$, it suffices to 
show that $\sigma_1$ is continuous
at $\delta=0$. 
Recall that $\sigma_1$ is defined as
\begin{equation} \label{eq:sigmaOneDef}
  \sigma_1(\delta)\coloneqq
  \sup_\mathcal{U}
  \max_{p\in \mathcal{P}^{(1)}_\mathcal{U}(\delta)}
  I(X_1,X_2;Y|U).
\end{equation}
Since in (\ref{eq:sigmaOneDef}), the supremum is over \emph{all} finite sets 
$\mathcal{U}$, it is difficult to find an upper
bound for $\sigma_1(\delta)$ near $\delta=0$ directly. 
Instead we first show that it is possible to assume
that $\mathcal{U}$ has at most two elements. 
\begin{lem}[Cardinality of $\mathcal{U}$] 
\label{lem:cardinality}
In the definition of $\sigma_1(\delta)$, 
it suffices to calculate the supremum over all sets $\mathcal{U}$
with $|\mathcal{U}|\leq 2$. 
\end{lem}

From Lemma \ref{lem:cardinality}, using standard tools,
such as Pinsker's inequality \cite[Lemma 17.3.3]{CoverThomas} and the 
$L_1$ bound on entropy \cite[Lemma 11.6.1]{CoverThomas}, 
the continuity of $\sigma_1$ at $\delta=0$ follows.\footnote{By 
Lemma \ref{lem:concavity}, $\sigma_1$ is concave 
on $\mathbb{R}_{\geq 0}$; thus, it is continuous 
on $\mathbb{R}_{>0}$.}
\begin{lem}[Continuity of $\sigma_1$] \label{lem:continuityFn}
The function $\sigma_1$ is continuous on $\mathbb{R}_{\geq 0}$.
\end{lem}

\section{Continuity of Sum-Capacity:
Arbitrary $\mathbf{C}_\mathrm{in}$}
\label{sec:arbitraryCF}

In this section, we study the continuity of
$C_\mathrm{sum}(\mathbf{C}_\mathrm{in},\mathbf{C}_\mathrm{out})$
over $\mathbb{R}^2_{\geq 0}\times\mathbb{R}^2_{\geq 0}$
with the aim of proving Theorem
\ref{thm:limitedCinContinuity}. 

Fix 
$(\mathbf{C}_\mathrm{in},\mathbf{C}_\mathrm{out})$.
For arbitrary 
$(\mathbf{\tilde{C}}_\mathrm{in},\mathbf{\tilde{C}}_\mathrm{out})$,
the triangle inequality implies
\begin{align}
  \MoveEqLeft
  \big|
  C_\mathrm{sum}(\mathbf{\tilde{C}}_\mathrm{in},\mathbf{\tilde{C}}_\mathrm{out})
  -C_\mathrm{sum}(\mathbf{C}_\mathrm{in},\mathbf{C}_\mathrm{out})\big|\notag\\
  &\leq \big|
  C_\mathrm{sum}(\mathbf{\tilde{C}}_\mathrm{in},\mathbf{\tilde{C}}_\mathrm{out})
  -C_\mathrm{sum}(\mathbf{C}_\mathrm{in},\mathbf{\tilde{C}}_\mathrm{out})
  \big|+ \big|
  C_\mathrm{sum}(\mathbf{C}_\mathrm{in},\mathbf{\tilde{C}}_\mathrm{out})
  -C_\mathrm{sum}(\mathbf{C}_\mathrm{in},\mathbf{C}_\mathrm{out})\big|
  \label{eq:triangleIneqArbitCin}
\end{align}
We study this bound in the limit
$(\mathbf{\tilde{C}}_\mathrm{in},\mathbf{\tilde{C}}_\mathrm{out})
\rightarrow
(\mathbf{C}_\mathrm{in},\mathbf{C}_\mathrm{out})$. 
We begin by considering the first term in (\ref{eq:triangleIneqArbitCin}).

\begin{lem}[Continuity of Sum-Capacity in $\mathbf{C}_\mathrm{in}$]
\label{lem:continuityCin}
There exists a function
\begin{equation*}
  \Delta\colon\mathbb{R}^2_{\geq 0}\times\mathbb{R}^2_{\geq 0}
  \rightarrow\mathbb{R}_{\geq 0}
\end{equation*}
that satisfies 
\begin{equation*}
  \lim_{\mathbf{\tilde{C}}_\mathrm{in}\rightarrow\mathbf{C}_\mathrm{in}}
  \Delta(\mathbf{C}_\mathrm{in},\mathbf{\tilde{C}}_\mathrm{in})=0,
\end{equation*}
and for any finite alphabet MAC and 
$(\mathbf{C}_\mathrm{in},\mathbf{\tilde{C}}_\mathrm{in},\mathbf{C}_\mathrm{out})
\in\mathbb{R}^2_{\geq 0}\times\mathbb{R}^2_{\geq 0}\times\mathbb{R}^2_{\geq 0}$,
we have
\begin{equation*}
  \big|C_\mathrm{sum}(\mathbf{C}_\mathrm{in},\mathbf{C}_\mathrm{out})
  -C_\mathrm{sum}(\mathbf{\tilde{C}}_\mathrm{in},\mathbf{C}_\mathrm{out})\big|
  \leq \Delta(\mathbf{C}_\mathrm{in},\mathbf{\tilde{C}}_\mathrm{in}).
\end{equation*}
\end{lem}

Applying Lemma \ref{lem:continuityCin} to (\ref{eq:triangleIneqArbitCin}), we get
\begin{equation*}
  \big|C_\mathrm{sum}(\mathbf{\tilde{C}}_\mathrm{in},\mathbf{\tilde{C}}_\mathrm{out})
  -C_\mathrm{sum}(\mathbf{C}_\mathrm{in},\mathbf{C}_\mathrm{out})\big|
  \leq \Delta(\mathbf{C}_\mathrm{in},\mathbf{\tilde{C}}_\mathrm{in})
  +\big|
  C_\mathrm{sum}(\mathbf{C}_\mathrm{in},\mathbf{\tilde{C}}_\mathrm{out})
  -C_\mathrm{sum}(\mathbf{C}_\mathrm{in},\mathbf{C}_\mathrm{out})\big|.
\end{equation*}
Thus to calculate the limit 
$(\mathbf{\tilde{C}}_\mathrm{in},\mathbf{\tilde{C}}_\mathrm{out})
\rightarrow (\mathbf{C}_\mathrm{in},\mathbf{C}_\mathrm{out})$, 
it suffices to consider 
\begin{equation*}
  \lim_{\mathbf{\tilde{C}}_\mathrm{out}
  \rightarrow \mathbf{C}_\mathrm{out}}
  \big|C_\mathrm{sum}(\mathbf{C}_\mathrm{in},\mathbf{\tilde{C}}_\mathrm{out})
  -C_\mathrm{sum}(\mathbf{C}_\mathrm{in},\mathbf{C}_\mathrm{out})\big|.
\end{equation*}
This is done in the next lemma.

\begin{lem}[Continuity of Sum-Capacity in $\mathbf{C}_\mathrm{out}$]
\label{lem:continuityCout}
For any finite alphabet MAC and 
$(\mathbf{C}_\mathrm{in},\mathbf{C}_\mathrm{out})
\in\mathbb{R}^2_{\geq 0}\times\mathbb{R}^2_{\geq 0}$,
proving that
\begin{equation*}
  \lim_{\mathbf{\tilde{C}}_\mathrm{out}\rightarrow\mathbf{C}_\mathrm{out}} 
  C_\mathrm{sum}(\mathbf{C}_\mathrm{in},\mathbf{\tilde{C}}_\mathrm{out})
  =C_\mathrm{sum}(\mathbf{C}_\mathrm{in},\mathbf{C}_\mathrm{out}).
\end{equation*}
is equivalent to showing that 
for all  
$(\mathbf{C}_\mathrm{in},\mathbf{C}_\mathrm{out})
\in\mathbb{R}^2_{>0}\times\mathbb{R}^2_{>0}$, we have
\begin{align*}
  \lim_{\tilde{C}_\mathrm{out}^1\rightarrow 0^+}
  C_\mathrm{sum}\big(\mathbf{C}_\mathrm{in},
  (\tilde{C}_\mathrm{out}^1,C_\mathrm{out}^2)\big)
  &=C_\mathrm{sum}\big(\mathbf{C}_\mathrm{in},
  (0,C_\mathrm{out}^2)\big)\\
  \lim_{\tilde{C}_\mathrm{out}^2\rightarrow 0^+}
  C_\mathrm{sum}\big(\mathbf{C}_\mathrm{in},
  (C_\mathrm{out}^1,\tilde{C}_\mathrm{out}^2)\big)
  &=C_\mathrm{sum}\big(\mathbf{C}_\mathrm{in},
  (C_\mathrm{out}^1,0)\big).
\end{align*}
\end{lem}

\section{Proofs}  \label{sec:proofs}

In this section, we begin with 
the proof of Corollary \ref{cor:discontinuity}.
We then provide detailed proofs of the 
lemmas appearing in Sections \ref{sec:fullKnowledgeCF} 
and \ref{sec:arbitraryCF}.

\subsection{Proof of Corollary \ref{cor:discontinuity}
(Necessary and Sufficient Condition for Discontinuity
of Maximal-Error Sum-Capacity)}

If (\ref{eq:avgBiggerThanMax}) holds, then by 
Proposition \ref{prop:discontinuity}, 
$C_\mathrm{sum,max}(\mathbf{C}_\mathrm{in},
\mathbf{C}_\mathrm{out})$
is not continuous at 
$\mathbf{C}_\mathrm{out}=\mathbf{0}$. Here we prove the
reverse direction. To this end, we show that if 
\begin{equation}  \label{eq:avgEqualsMax}
  C_\mathrm{sum,avg}(\mathbf{0},\mathbf{0})
  =C_\mathrm{sum,max}(\mathbf{0},\mathbf{0}),
\end{equation}
then $C_\mathrm{sum,max}(\mathbf{C}_\mathrm{in},
\mathbf{C}_\mathrm{out})$
is continuous at 
$\mathbf{C}_\mathrm{out}=\mathbf{0}$.

We begin by defining the function 
$f\colon\mathbb{R}_{\geq 0}\rightarrow
\mathbb{R}_{\geq 0}$ as 
\begin{equation*}
  f(C_\mathrm{out}) \coloneqq
  C_\mathrm{sum,max}(\mathbf{C}_\mathrm{in}^*,
  (C_\mathrm{out},C_\mathrm{out}))
\end{equation*}
Then by \cite[Theorem 1]{reliability} and 
(\ref{eq:avgEqualsMax}), for all 
$C_\mathrm{out}\geq 0$, we have
\begin{equation*}
  f(C_\mathrm{out})=
  C_\mathrm{sum,avg}(\mathbf{C}_\mathrm{in}^*,
  (C_\mathrm{out},C_\mathrm{out})).
\end{equation*}
For each $\mathbf{C}_\mathrm{out}\in\mathbb{R}^2_{\geq 0}$,
let $C_\mathrm{out}^*\coloneqq\max\{C_\mathrm{out}^1,C_\mathrm{out}^2\}$.
Then for any $\mathbf{C}_\mathrm{in}\in\mathbb{R}^2_{\geq 0}$,
\begin{equation*}
  f(0)\leq C_\mathrm{sum,max}(\mathbf{C}_\mathrm{in},
  \mathbf{C}_\mathrm{out})\leq f(C_\mathrm{out}^*).
\end{equation*}
If we now let $\mathbf{C}_\mathrm{out}\rightarrow\mathbf{0}$
and apply Theorem \ref{thm:CsumAvgContinuity}, the continuity of 
$C_\mathrm{sum,max}(\mathbf{C}_\mathrm{in},
\mathbf{C}_\mathrm{out})$
at $\mathbf{C}_\mathrm{out}=\mathbf{0}$ follows.  

\subsection{Proof of Lemma \ref{lem:superadditivity}}

By the definition of $\sigma_n(\delta)$, for all $\epsilon>0$, 
there exist finite alphabets 
$\mathcal{U}_0$ and $\mathcal{U}_1$ 
and distributions $p_n\in \mathcal{P}^{(n)}_{\mathcal{U}_0}(\delta)$
and $p_m\in \mathcal{P}^{(m)}_{\mathcal{U}_1}(\delta)$ such that
\begin{align*}
  I_n(X_1^n,X_2^n;Y^n|U_0) &\geq n\sigma_n(\delta)-n\epsilon\\
  I_m(X_1^m,X_2^m;Y^m|U_1) &\geq m\sigma_m(\delta)-m\epsilon.
\end{align*}
Consider the distribution
\begin{equation*}
  p_{n+m}(u_0,u_1,x_1^{n+m},x_2^{n+m})
  =p_n(u_0,x_1^n,x_2^n)p_m(u_1,x_1^{n+1:n+m},x_2^{n+1:n+m}).
\end{equation*}
Let $\mathcal{U}\coloneqq\mathcal{U}_0\times\mathcal{U}_1$.
Then it is straightforward to show that
that $p_{n+m}\in\mathcal{P}_\mathcal{U}^{(n+m)}(\delta)$, and 
\begin{equation*}
  I_{n+m}(X_1^{n+m},X_2^{n+m};Y^{n+m}|U_0,U_1)
  \geq n\sigma_n(\delta)+m\sigma_m(\delta)-(n+m)\epsilon,
\end{equation*}
which implies the desired result. 

\subsection{Proof of Lemma \ref{lem:CsumAvgBounds}}

We first prove the lower bound. For $i\in\{1,2\}$,
choose $C_{id}$ such that 
\begin{equation*}
0\leq C_{id}\leq C_\mathrm{out}^i.
\end{equation*}
Let $p(u,x_1,x_2)$ be any distribution satisfying 
\begin{equation*}
  I(X_1;X_2|U)=C_{1d}+C_{2d}.
\end{equation*}
Then \cite[Corollary 8]{kUserMAC} implies that 
\begin{equation*}
  C_\mathrm{sum}(\mathbf{C}_\mathrm{in}^*,
  \mathbf{C}_\mathrm{out})\geq
  I(X_1,X_2;Y|U)-\min\{C_{1d},C_{2d}\}.
\end{equation*}
Applying the same corollary to the
MAC
\begin{equation*}
  p(y^n|x_1^n,x_2^n)=\prod_{t\in [n]}
  p(y_t|x_{1t},x_{2t}),
\end{equation*}
proves our lower bound.

For the upper bound, consider a sequence of 
$(2^{nR_1},2^{nR_2},n)$-codes for the MAC 
with a $(\mathbf{C}_\mathrm{in}^*,\mathbf{C}_\mathrm{out})$-CF.
By the data processing inequality,
\begin{equation*}
  I(X_1^n;X_2^n)\leq n(C_\mathrm{out}^1+C_\mathrm{out}^2).
\end{equation*}
In addition, from Fano's inequality it follows that 
there exists a sequence $(\epsilon_n)_{n=1}^\infty$
such that 
\begin{equation*}
  H(W_1,W_2|Y^n)\leq n\epsilon_n,
\end{equation*}
and $\epsilon_n\rightarrow 0$ as $n\rightarrow\infty$. 
We have
\begin{align*}
  n(R_1+R_2) &= H(W_1,W_2)\\
  &= I(W_1,W_2;Y^n)+H(W_1,W_2|Y^n)\\
  &= I(X_1^n,X_2^n;Y^n)+n\epsilon_n\\
  &\leq n\sigma(C_\mathrm{out}^1+C_\mathrm{out}^2)
  +n\epsilon_n.
\end{align*}
Dividing by $n$ and taking the limit $n\rightarrow\infty$
completes the proof. 

\subsection{Proof of Lemma \ref{lem:concavity} (Concavity of $\sigma_n$)}

It suffices to prove the result for $n=1$. 
We apply the technique from \cite{CoverElGamalSalehi}.
Note that
\begin{equation*}
  \sigma_1(\delta)= \sup_\mathcal{U}
  \max_{p\in \mathcal{P}^{(1)}_\mathcal{U}(\delta)}
  I(X_1,X_2;Y|U). 
\end{equation*}
Fix $a,b\geq 0$, $\lambda\in (0,1)$, and $\epsilon>0$. 
Then there exist finite sets $\mathcal{U}_0$ and $\mathcal{U}_1$
and distributions
$p_0\in \mathcal{P}^{(1)}_{\mathcal{U}_0}(a)$
and $p_1\in \mathcal{P}^{(1)}_{\mathcal{U}_1}(b)$ 
satisfying 
\begin{align*}
  I_0(X_1,X_2;Y|U_0) &\geq \sigma_1(a)-\epsilon\\
  I_1(X_1,X_2;Y|U_1) &\geq \sigma_1(b)-\epsilon,
\end{align*}
respectively. Define the alphabet $\mathcal{V}$ as
\begin{equation*}
  \mathcal{V}\coloneqq \{0\}\times\mathcal{U}_0
  \cup \{1\}\times\mathcal{U}_1.
\end{equation*}
We denote an element of $\mathcal{V}$ by
$v=(v_1,v_2)$. 
Define the distribution $p_\lambda(v,x_1,x_2)$ as
\begin{equation*}
  p_\lambda(v,x_1,x_2)
  =p_\lambda(v_1)
  p_{v_1}(v_2,x_1,x_2),
\end{equation*}
where 
\begin{equation*}
  p_\lambda(v_1)
  =\begin{cases}
  1-\lambda &\text{if }v_1=0\\
  \lambda &\text{if }v_1=1.
  \end{cases}
\end{equation*}
Then
\begin{align*}
  I_\lambda(X_1;X_2|V)
  &= I_\lambda(X_1,X_2|V_1,V_2)\\
  &= (1-\lambda)I(X_1;X_2|V_1=0,V_2)
  +\lambda I(X_1;X_2|V_1=1,V_2)\\
  &= (1-\lambda)I_0(X_1;X_2|U_0)
  +\lambda I_1(X_1;X_2|U_1)\\
  &\leq (1-\lambda)a+\lambda b,
\end{align*}
which implies 
$p_\lambda\in\mathcal{P}^{(1)}_\mathcal{V}
((1-\lambda)a+\lambda b)$. Similarly,
\begin{align*}
  I_\lambda(X_1,X_2;Y|V)
  &= I_\lambda(X_1,X_2;Y|V_1,V_2)\\
  &= (1-\lambda)I(X_1,X_2;Y|V_1=0,V_2)
  +\lambda I(X_1,X_2;Y|V_1=1,V_2)\\
  &= (1-\lambda)I_0(X_1,X_2;Y|U_0)
  +\lambda I_1(X_1,X_2;Y|U_1)\\
  &\geq (1-\lambda)\sigma_1(a)+\lambda \sigma_1(b)-\epsilon.
\end{align*}
Therefore, 
\begin{equation*}
  \sigma_1\big((1-\lambda)a+\lambda b\big)\geq 
  (1-\lambda)\sigma_1(a)+\lambda \sigma_1(b)-\epsilon.
\end{equation*}
The result now follows from the fact that the above
equation holds for all $\epsilon >0$. 

\subsection{Proof of Lemma \ref{lem:cardinality}}

Let $\mathcal{U}$ be some 
finite set and let $p^*\in\mathcal{P}^{(1)}_\mathcal{U}(\delta)$
be a distribution that achieves
\begin{equation*}
  \max_{p\in \mathcal{P}^{(1)}_\mathcal{U}(\delta)}
  I(X_1,X_2;Y|U).
\end{equation*}
Let $\mathcal{Q}\subseteq\mathbb{R}^{|\mathcal{U}|}$ denote the set of all
vectors $(q(u))_{u\in\mathcal{U}}$ that satisfy
\begin{align}
  &q(u)\geq 0\text{ for all }u\in\mathcal{U}\notag\\
  &\sum_{u\in\mathcal{U}}q(u)=1\notag\\
  &\sum_{u\in\mathcal{U}}q(u)I^*(X_1;X_2|U=u)
  =I^*(X_1;X_2|U),\label{eq:qMutualInfo}
\end{align}
where in (\ref{eq:qMutualInfo}), $I^*(X_1;X_2|U=u)$
and $I^*(X_1;X_2|U)$ are calculated according to
$p^*(x_1,x_2|u)$ and $p^*(u,x_1,x_2)$, respectively. 
Consider the mapping $F\colon\mathcal{Q}\rightarrow\mathbb{R}_{\geq 0}$ 
defined by
\begin{equation} \label{eq:mappingFdef}
  F[q]\coloneqq 
  \sum_{u\in\mathcal{U}}q(u)I^*(X_1,X_2;Y|U=u),
\end{equation}
where $I^*(X_1,X_2;Y|U=u)$ is calculated with respect to 
$p^*(x_1,x_2|u)p(y|x_1,x_2)$. Note that since 
$p^*(u)\in\mathcal{Q}$, we have
\begin{equation*}
  \max_{q\in\mathcal{Q}}F[q]= 
  \max_{p\in \mathcal{P}^{(1)}_\mathcal{U}(\delta)}
  I(X_1,X_2;Y|U).
\end{equation*}
Thus it suffices to find $q^*\in\mathcal{Q}$ which has at most
two non-zero components and at which $F$ 
obtains its maximal value.

Since $F$ is linear in $q$, it is convex on $\mathcal{Q}$,
which is a bounded convex polyhedron in 
$\mathbb{R}^{|\mathcal{U}|}$. Thus there exists an extreme point of
$\mathcal{Q}$, say $q^*\in\mathcal{Q}$, at which $F$ obtains its
maximum. Since $q^*$ is an extreme point, if we apply 
\cite[p. 50, Theorem 2.3]{BertsimasTsitsiklis} to the definition 
of $\mathcal{Q}$, we see that
we must have $q^*(u)=0$ for at least 
\begin{equation*}
  |\mathcal{U}|-2
\end{equation*}
values of $u$. This completes the proof. 

\subsection{Proof of Lemma \ref{lem:continuityFn} (Continuity of $\sigma_1$)}

By Lemma \ref{lem:cardinality}, without loss of generality, we can
set $\mathcal{U}\coloneqq\{a,b\}$. 
For all $\delta\geq 0$, we have
\begin{equation*}
  \sigma_1(\delta)= 
  \max_{p\in \mathcal{P}^{(1)}_\mathcal{U}(\delta)}
  I(X_1,X_2;Y|U). 
\end{equation*}
Fix $\delta>0$. Let $p^*(u,x_1,x_2)$ be a distribution 
in $\mathcal{P}_\mathcal{U}^{(1)}(\delta)$ achieving
the maximum above, and define
\begin{equation*}
  p^*_\mathrm{ind}(x_1,x_2|u)\coloneqq 
  p^*(x_1|u)p^*(x_2|u).
\end{equation*}
Since
\begin{equation*}
  \sum_{u\in\mathcal{U}}
  p^*(u)D\big(p^*(x_1,x_2|u)\|p^*_\mathrm{ind}(x_1,x_2|u)\big)
  = I^*(X_1;X_2|U)\leq \delta,
\end{equation*}
by \cite[Lemma 11.6.1]{CoverThomas},
\begin{equation} \label{eq:L1bound}
  \sum_{u\in\mathcal{U}}p^*(u)
  \big\| p^*(x_1,x_2|u)-p^*_\mathrm{ind}(x_1,x_2|u)\big\|^2_{L^1}
  \leq 2\delta\ln 2.
\end{equation}
In addition,  
\begin{align}
  \MoveEqLeft
  \sum_{u\in\mathcal{U}}p^*(u)
  \big\|p^*(y|u)-p^*_\mathrm{ind}(y|u)\big\|_{L^1}\notag\\
  &\leq \sum_{u\in\mathcal{U}}p^*(u)\sum_{x_1,x_2}p(y|x_1,x_2)\big|
  p^*(x_1,x_2|u)-p^*_\mathrm{ind}(x_1,x_2|u)\big|\notag\\
  &\leq \sum_{u\in\mathcal{U}}p^*(u)
  \big\| p^*(x_1,x_2|u)-p^*_\mathrm{ind}(x_1,x_2|u)\big\|_{L^1}
  \notag\\
  &\leq \sqrt{2\delta\ln 2},\label{eq:twoDelta}
\end{align}
where (\ref{eq:twoDelta}) follows from (\ref{eq:L1bound}) and
the Cauchy-Schwarz inequality. 
Define the subset $\mathcal{U}_0\subseteq\mathcal{U}$ as
\begin{equation*}
  \mathcal{U}_0=\Big\{u\in\mathcal{U}:
  \big\|p^*(y|u)-p^*_\mathrm{ind}(y|u)\big\|_{L^1}
  \leq 1/2\Big\}.
\end{equation*}
Clearly, by (\ref{eq:twoDelta}),
\begin{equation} \label{eq:PrNotInU0}
  \sum_{u\notin\mathcal{U}_0}p^*(u)
  \leq 2\sqrt{2\delta\ln 2}.
\end{equation}
Thus
\begin{align}
  \MoveEqLeft
  \big|H^*(Y|U)-H^*_\mathrm{ind}(Y|U)\big|\notag\\
  &\leq \sum_{u\in\mathcal{U}}p^*(u)
  \big|H^*(Y|U=u)-H^*_\mathrm{ind}(Y|U=u)\big|\notag\\
  &\overset{(a)}{\leq} 2\sqrt{2\delta\ln 2}\log |\mathcal{Y}|\notag\\
  &\phantom{\leq}
  -\sum_{u\in\mathcal{U}_0}p^*(u)
  \big\|p^*(y|u)-p^*_\mathrm{ind}(y|u)\big\|_{L^1}
  \log\frac{\big\|p^*(y|u)-p^*_\mathrm{ind}(y|u)\big\|_{L^1}}
  {|\mathcal{Y}|}\notag\\
  &\overset{(b)}{\leq} 2\sqrt{2\delta\ln 2}\log |\mathcal{Y}|-\sqrt{2\delta\ln 2}\log\Big(
  \frac{1}{|\mathcal{Y}|}\sqrt{2\delta\ln 2}\Big)\notag\\
  &=\sqrt{2\delta\ln 2}\log\frac{|\mathcal{Y}|^3}{\sqrt{2\delta\ln 2}},\label{eq:EntropyBoundDifference}
\end{align}
where $(a)$ follows from (\ref{eq:PrNotInU0}) and \cite[Theorem 17.3.3]{CoverThomas},
and $(b)$ follows from (\ref{eq:twoDelta}) and the fact that the mapping 
$t\mapsto -t\log (t/|\mathcal{Y}|)$ is concave on its domain and increasing for sufficiently
small $t$.\footnote{Precisely, the mapping $t\mapsto -t\log (t/|\mathcal{Y}|)$
is increasing on $(0,|\mathcal{Y}|/e)$.} In addition, by (\ref{eq:twoDelta}),
\begin{align}
  \MoveEqLeft
  \big|H^*(Y|U,X_1,X_2)-H^*_\mathrm{ind}(Y|U,X_1,X_2)\big|\notag\\
  &\leq \sum_{u,x_1,x_2}
  \big|p^*(u,x_1,x_2)-p^*_\mathrm{ind}(u,x_1,x_2)\big|H(Y|X_1=x_1,X_2=x_2)
  \notag\\
  &\leq \max_{x_1,x_2}H(Y|X_1=x_1,X_2=x_2)\cdot
  \sum_{u\in\mathcal{U}_0}p^*(u)
  \big\|p^*(x_1,x_2|u)-p^*_\mathrm{ind}(x_1,x_2|u)\big\|_{L^1}
  \notag\\
  &\leq \Big(\log|\mathcal{Y}|\Big)\sqrt{2\delta\ln 2}.
  \label{eq:logYTwoDelta}
\end{align}
Thus by (\ref{eq:EntropyBoundDifference}) and
(\ref{eq:logYTwoDelta}),
\begin{align*}
  \sigma_1(\delta) &= I^*(X_1,X_2;Y|U)
  =H^*(Y|U)-H^*(Y|U,X_1,X_2)\\
  &\leq \big|H^*(Y|U)-H^*_\mathrm{ind}(Y|U)\big|
  +\big|H^*(Y|U,X_1,X_2)-H^*_\mathrm{ind}(Y|U,X_1,X_2)\big|\\
  &\phantom{=}
  +I^*_\mathrm{ind}(X_1,X_2;Y|U)\\
  &\leq \sqrt{2\delta\ln 2}\log\frac{|\mathcal{Y}|^3}{\sqrt{2\delta\ln 2}}
  +\Big(\log|\mathcal{Y}|\Big)\sqrt{2\delta\ln 2}
  +\sigma_1(0).
\end{align*}
Since $\sigma_1(0)\leq \sigma_1(\delta)$ for all $\delta\geq 0$, 
the continuity of $\sigma_1$ at $\delta=0^+$ follows. 

\subsection{Proof of Lemma \ref{lem:Dueck} (Dueck's Lemma)} 
\label{subsec:Dueck}

If for all $t\in [n]$, we have
\begin{equation*}
  I(X_{1t};X_{2t}|U)\leq \epsilon,
\end{equation*}
then we define $T\coloneqq \varnothing$. Otherwise,
there exists $t_1\in [n]$ such that
\begin{equation} \label{eq:singleLetterBoundMI}
  I(X_{1t_1};X_{2t_1}|U)>\epsilon.
\end{equation}
Let $S_1\coloneqq [n]\setminus\{t_1\}$. Then 
\begin{align*}
  I(X_1^n;X_2^n|U) &= I(X_1^n;X_{2t_1}|U)+I(X_1^n;X_2^{S_1}|U,X_{2t_1})\\
  &= I(X_{1t_1};X_{2t_1}|U)+I(X_1^{S_1};X_{2t_1}|U,X_{1t_1})\\
  &\phantom{=}+I(X_{1t_1};X_{2}^{S_1}|U,X_{2t_1})+I(X_1^{S_1};X_2^{S_1}|U,X_{1t_1},X_{2t_1})\\
  &\geq I(X_{1t_1};X_{2t_1}|U)+I(X_1^{S_1};X_2^{S_1}|U,X_{1t_1},X_{2t_1}).
\end{align*}
Since $I(X_1^n;X_2^n|U)\leq n\delta$, using (\ref{eq:singleLetterBoundMI}),
we get
\begin{equation*}
  I(X_1^{S_1};X_2^{S_1}|U,X_{1t_1},X_{2t_1})
  \leq n\delta-\epsilon.
\end{equation*}
Now if for all $t\in S_1$, 
\begin{equation*}
  I(X_{1t};X_{2t}|U,X_{1t_1},X_{2t_1})\leq\epsilon,
\end{equation*}
then we define $T\coloneqq \{t_1\}$. Otherwise, there exists $t_2\in [n]$ such that 
\begin{equation*}
  I(X_{1t_2};X_{2t_2}|U,X_{1t_1},X_{2t_1})>\epsilon.
\end{equation*}
Similar to the above argument, if we define
$S_2\coloneqq [n]\setminus\{t_1,t_2\}$,  then
\begin{equation*} I(X_1^{S_2};X_2^{S_2}|U,X_{1t_1},X_{1t_2},X_{2t_1},X_{2t_2})
  \leq n\delta-2\epsilon.
\end{equation*}
If we continue this process, we eventually get a set  $T\coloneqq \{t_1,\dots,t_k\}$ such that
\begin{equation} \label{eq:SkTeps}
  I(X_1^{T^c};X_2^{T^c}|U,X_{1}^T,X_{2}^T)
  \leq n\delta-|T|\epsilon,
\end{equation}
and for all $t\in S_k\coloneqq T^c$,
\begin{equation*}
  I(X_{1t};X_{2t}|U,X_{1}^T,X_{2}^T)\leq\epsilon.
\end{equation*}
In addition, from (\ref{eq:SkTeps}) it follows that
\begin{equation}
  |T|\leq \frac{n\delta}{\epsilon}.
\end{equation}

\subsection{Proof of Corollary \ref{cor:Dueck}} \label{subsec:CorDueck}

Fix a positive integer $n$. By Lemma \ref{lem:cardinality}, we
can set $\mathcal{U}\coloneqq\{a,b\}$. 
From Lemma \ref{lem:Dueck}, it follows that there exists
a set $T\subseteq [n]$ such that
\begin{equation} \label{eq:boundCardinalityT}
  0\leq |T|\leq \frac{n\delta}{\epsilon},
\end{equation}
and 
\begin{equation*}
  \forall\: t\notin T\colon
  I(X_{1t};X_{2t}|U,X_1^T,X_2^T)\leq\epsilon.
\end{equation*}
Thus
\begin{align}
  I(X_1^n,X_2^n;Y^n|U)
  &= I(X_1^T,X_2^T;Y^n|U)+I(X_1^{T^c},X_2^{T^c};Y^n|U,X_1^T,X_2^T)
  \notag\\
  &\leq |T|\log |\mathcal{X}_1||\mathcal{X}_2|
  +I(X_1^{T^c},X_2^{T^c};Y^n|U,X_1^T,X_2^T).
  \label{eq:boundInputOutputMI}
\end{align}
We further bound the second term on the right hand side by 
\begin{align}
  \MoveEqLeft
  I(X_1^{T^c},X_2^{T^c};Y^n|U,X_1^T,X_2^T)\notag\\
  &= I(X_1^{T^c},X_2^{T^c};Y^{T^c}|U,X_1^T,X_2^T)
  +I(X_1^{T^c},X_2^{T^c};Y^T|U,X_1^T,X_2^T,Y^{T^c})\notag\\
  &\leq \sum_{t\notin T}
  I(X_{1t},X_{2t};Y_t|U,X_1^T,X_2^T)\notag\\
  &\leq n\max_{p\in\mathcal{P}^{(1)}_{\mathcal{V}}(\epsilon)}
  I(X_1,X_2;Y|V)\leq n\sigma_1(\epsilon),
  \label{eq:TcYBoundConditionedT}
\end{align}
where 
\begin{equation*}
  \mathcal{V}\coloneqq \mathcal{U}\times\mathcal{X}_1^{|T|}\times\mathcal{X}_2^{|T|}.
\end{equation*}
Therefore, by (\ref{eq:boundCardinalityT}), (\ref{eq:boundInputOutputMI}), 
and (\ref{eq:TcYBoundConditionedT}),
\begin{equation*}
  \frac{1}{n}I(X_1^n,X_2^n;Y^n|U)\leq
  \frac{\delta}{\epsilon}\log 
  |\mathcal{X}_1||\mathcal{X}_2|
  +\sigma_1(\epsilon),
\end{equation*}
which completes the proof.

\subsection{Proof of Lemma \ref{lem:continuityCin}
(Continuity of Sum-Capacity in $\mathbf{C}_\mathrm{in}$)}
\label{subsec:continuityCin}

Fix $\mathbf{C}_\mathrm{out}\in\mathbb{R}^2_{\geq 0}$. 
Define the functions $f,g\colon\mathbb{R}^2_{\geq 0}\rightarrow\mathbb{R}$
as 
\begin{align*}
  f(\mathbf{C}_\mathrm{in}) &\coloneqq C_\mathrm{sum}(\mathbf{C}_\mathrm{in},\mathbf{C}_\mathrm{out})
  -C_\mathrm{sum}(\mathbf{C}_\mathrm{in},\mathbf{0})
  =C_\mathrm{sum}(\mathbf{C}_\mathrm{in},\mathbf{C}_\mathrm{out})
  -C_\mathrm{sum}(\mathbf{0},\mathbf{0})\\
  g(\mathbf{C}_\mathrm{in}) &\coloneqq
  C_\mathrm{in}^1+C_\mathrm{in}^2-g(\mathbf{C}_\mathrm{in}).
\end{align*}
Note that since $f$ is concave, $g$ is convex. Thus 
for all $\lambda\in [0,1]$ and all 
$(\mathbf{C}_\mathrm{in},\mathbf{\tilde{C}}_\mathrm{in})$,
\begin{equation*}
  g(\lambda\mathbf{C}_\mathrm{in}
  +(1-\lambda)\mathbf{\tilde{C}}_\mathrm{in})
  \leq \lambda g(\mathbf{C}_\mathrm{in})
  +(1-\lambda)g(\mathbf{\tilde{C}}_\mathrm{in}).
\end{equation*}
Since $g(\mathbf{0})=0$, setting 
$\mathbf{\tilde{C}}_\mathrm{in}=\mathbf{0}$
gives 
\begin{equation*}
  g(\lambda\mathbf{C}_\mathrm{in})
  \leq \lambda g(\mathbf{C}_\mathrm{in})
\end{equation*}
Note that by \cite[Proposition 6]{kUserMAC}, $g$ is nonnegative. Thus
\begin{equation*}
  g(\lambda\mathbf{C}_\mathrm{in})
  \leq g(\mathbf{C}_\mathrm{in}),
\end{equation*}
which when written in terms of $f$, is equivalent
to
\begin{equation} \label{eq:sumCapacityBoundCin}
  f(\mathbf{C}_\mathrm{in})
  -f(\lambda\mathbf{C}_\mathrm{in})
  \leq (1-\lambda)(C_\mathrm{in}^1+C_\mathrm{in}^2).
\end{equation}

Consider 
$\mathbf{C}_\mathrm{in},\mathbf{\tilde{C}}_\mathrm{in}
\in\mathbb{R}^2_{\geq 0}$. Define the pairs 
$\mathbf{\underaccent{\bar} C}_\mathrm{in},\mathbf{\bar C}_\mathrm{in}
\in\mathbb{R}^2_{\geq 0}$ as
\begin{align*}
  &\forall\:i\in\{1,2\}\colon
  \underaccent{\bar}{C}_\mathrm{in}^i
  \coloneqq \min\{C_\mathrm{in}^i,\tilde{C}_\mathrm{in}^i\}\\
  &\forall\:i\in\{1,2\}\colon
  \bar{C}_\mathrm{in}^i
  \coloneqq \max\{C_\mathrm{in}^i,\tilde{C}_\mathrm{in}^i\}\\
\end{align*}
Next define 
$\lambda^*(\mathbf{C}_\mathrm{in},\mathbf{\tilde{C}}_\mathrm{in})\in [0,1]$ 
as\footnote{If for some $i\in\{1,2\}$, say $i=1$, 
$\bar{C}_\mathrm{in}^i=0$, set $\lambda^*\coloneqq
\min\{1,\underaccent{\bar}{C}_\mathrm{in}^2/\bar{C}_\mathrm{in}^2\}$.
If $\bar{C}_\mathrm{in}^1=\bar{C}_\mathrm{in}^2=0$, set 
$\lambda^*=1$. These definitions ensure the continuity of 
$\lambda^*$ in $(\mathbf{C}_\mathrm{in},\mathbf{\tilde{C}}_\mathrm{in})$.}
\begin{equation*}
  \lambda^*\coloneqq
  \min_{i\in\{1,2\}}\underaccent{\bar}{C}_\mathrm{in}^i/
  \bar{C}_\mathrm{in}^i.
\end{equation*}
Then
\begin{align}
  \big|f(\mathbf{C}_\mathrm{in})-f(\mathbf{\tilde C}_\mathrm{in})\big|
  &\leq \big|f(\mathbf{C}_\mathrm{in})-f(\mathbf{\underaccent{\bar} C}_\mathrm{in})\big|
  +\big|f(\mathbf{\underaccent{\bar} C}_\mathrm{in})-f(\mathbf{\tilde C}_\mathrm{in})\big|
  \label{eq:continuityCin1}\\
  &=f(\mathbf{C}_\mathrm{in})-f(\mathbf{\underaccent{\bar} C}_\mathrm{in})
  +f(\mathbf{\tilde C}_\mathrm{in})-f(\mathbf{\underaccent{\bar} C}_\mathrm{in})
  \label{eq:continuityCin2}\\
  &\leq f(\mathbf{C}_\mathrm{in})-f(\lambda^*\mathbf{C}_\mathrm{in})
  +f(\mathbf{\tilde C}_\mathrm{in})-f(\lambda^*\mathbf{\tilde C}_\mathrm{in})
  \label{eq:continuityCin3}\\
  &\leq (1-\lambda^*)(C_\mathrm{in}^1+C_\mathrm{in}^2)
  +(1-\lambda^*)(\tilde{C}_\mathrm{in}^1+\tilde{C}_\mathrm{in}^2),
  \label{eq:continuityCin4}
\end{align}
where (\ref{eq:continuityCin1}) follows from the triangle inequality,
(\ref{eq:continuityCin2}) follows from the definition of 
$\mathbf{\underaccent{\bar} C}_\mathrm{in}$, 
(\ref{eq:continuityCin3}) follows from the definition of 
$\lambda^*$, and (\ref{eq:continuityCin4}) follows from
(\ref{eq:sumCapacityBoundCin}). Finally, if we let 
$\mathbf{\tilde{C}}_\mathrm{in}\rightarrow\mathbf{C}_\mathrm{in}$
in (\ref{eq:continuityCin3}), we see that 
$f(\mathbf{\tilde{C}}_\mathrm{in})\rightarrow
f(\mathbf{C}_\mathrm{in})$, since 
\begin{equation*}
  \lim_{\mathbf{\tilde{C}}_\mathrm{in}\rightarrow\mathbf{C}_\mathrm{in}}
  \lambda^*(\mathbf{C}_\mathrm{in},\mathbf{\tilde{C}}_\mathrm{in})=1.
\end{equation*}

\subsection{Proof of Lemma \ref{lem:continuityCout}
(Continuity of Sum-Capacity in $\mathbf{C}_\mathrm{out}$)}
\label{subsec:continuityCout}

Recall that we only need to verify continuity on the boundary
of $\mathbb{R}^2_{\geq 0}\times\mathbb{R}^2_{\geq 0}$; namely,
the set of all points 
$(\mathbf{C}_\mathrm{in},\mathbf{C}_\mathrm{out})$ where at least
one of $C_\mathrm{in}^1$, $C_\mathrm{in}^2$, $C_\mathrm{out}^1$,
or $C_\mathrm{out}^2$ is zero. 

We first show that 
\begin{equation*}
  \lim_{\mathbf{\tilde{C}}_\mathrm{out}\rightarrow\mathbf{C}_\mathrm{out}} 
  C_\mathrm{sum}(\mathbf{C}_\mathrm{in},\mathbf{\tilde{C}}_\mathrm{out})
  =C_\mathrm{sum}(\mathbf{C}_\mathrm{in},\mathbf{C}_\mathrm{out}).
\end{equation*}
holds if $\mathbf{C}_\mathrm{out}=\mathbf{0}$, or if either
$C_\mathrm{in}^1=0$ or $C_\mathrm{in}^2=0$. For the case
$\mathbf{C}_\mathrm{out}=\mathbf{0}$, note that 
\begin{align*}
  C_\mathrm{sum}(\mathbf{C}_\mathrm{in},\mathbf{\tilde{C}}_\mathrm{out})
  -C_\mathrm{sum}(\mathbf{C}_\mathrm{in},\mathbf{0})
  &= C_\mathrm{sum}(\mathbf{C}_\mathrm{in},\mathbf{\tilde{C}}_\mathrm{out})
  -C_\mathrm{sum}(\mathbf{0},\mathbf{0})\\
  &\leq C_\mathrm{sum}(\mathbf{C}_\mathrm{in}^*,\mathbf{\tilde{C}}_\mathrm{out})
  -C_\mathrm{sum}(\mathbf{0},\mathbf{0}),
\end{align*}
which goes to zero as $\mathbf{\tilde{C}}_\mathrm{out}\rightarrow\mathbf{0}$
by Theorem \ref{thm:CsumAvgContinuity}.

Next suppose $C_\mathrm{in}^2=0$. In this case, we have
\begin{equation*}
  C_\mathrm{sum}\big((C_\mathrm{in}^1,0),\mathbf{\tilde{C}}_\mathrm{out}\big)
  = C_\mathrm{sum}\big((C_\mathrm{in}^1,0),(0,\tilde{C}_\mathrm{out}^2)\big).
\end{equation*}
Let $f\colon\mathbb{R}_{\geq 0}\rightarrow\mathbb{R}_{\geq 0}$
denote the function
\begin{equation*}
  f(C_\mathrm{out}^2)
  \coloneqq 
  C_\mathrm{sum}\big((C_\mathrm{in}^1,0),(0,C_\mathrm{out}^2)\big).
\end{equation*} 
Note that $f$ is continuous on $\mathbb{R}_{>0}$ since it is concave.
To prove the continuity of $f$ at $C_\mathrm{out}^2=0$, observe that
$f(C_\mathrm{out}^2)$ equals the sum-capacity of a MAC with 
a $(C_{12},0)$-conference \cite{WillemsMAC}, where 
\begin{equation*}
  C_{12}\coloneqq\min\{C_\mathrm{in}^1,C_\mathrm{out}^2\}.
\end{equation*}
From the capacity region given in \cite{WillemsMAC}, we have
\begin{equation*}
  f(C_\mathrm{out}^2)
  \leq f(0)+\min\{C_\mathrm{in}^1,C_\mathrm{out}^2\},
\end{equation*}
from which implies that $f$ is 
continuous at $C_\mathrm{out}^2=0$.
The case where $C_\mathrm{in}^1=0$ follows similarly. 

Finally, consider the case where $C_\mathrm{out}^2=0$, but 
$C_\mathrm{out}^1>0$. In this case, we apply
the next lemma for concave functions that are nondecreasing as
well. 
\begin{lem} \label{lem:concaveFunctions}
Let $f\colon\mathbb{R}_{\geq 0}\rightarrow\mathbb{R}_{\geq 0}$ be
concave and nondecreasing. Then if $|x-y|\leq \min\{x,y\}$,
\begin{equation*}
  \big|f(x)-f(y)\big|
  \leq f(|x-y|)-f(0).
\end{equation*}
\end{lem}

We have 
\begin{align*}
  \MoveEqLeft
  \big|C_\mathrm{sum}(\mathbf{C}_\mathrm{in},(\tilde{C}_\mathrm{out}^1,\tilde{C}_\mathrm{out}^2))
  -C_\mathrm{sum}(\mathbf{C}_\mathrm{in},(C_\mathrm{out}^1,0))\big|\\
  &\overset{(a)}{\leq}
  \big|C_\mathrm{sum}(\mathbf{C}_\mathrm{in},(\tilde{C}_\mathrm{out}^1,\tilde{C}_\mathrm{out}^2))
  -C_\mathrm{sum}(\mathbf{C}_\mathrm{in},(C_\mathrm{out}^1,\tilde{C}_\mathrm{out}^2))\big|
  +\big|C_\mathrm{sum}(\mathbf{C}_\mathrm{in},(C_\mathrm{out}^1,\tilde{C}_\mathrm{out}^2))
  -C_\mathrm{sum}(\mathbf{C}_\mathrm{in},(C_\mathrm{out}^1,0))\big|\\
  &\overset{(b)}{\leq}
  \big|C_\mathrm{sum}(\mathbf{C}_\mathrm{in},(|\tilde{C}_\mathrm{out}^1-C_\mathrm{out}^1|,\tilde{C}_\mathrm{out}^2))
  -C_\mathrm{sum}(\mathbf{C}_\mathrm{in},(0,\tilde{C}_\mathrm{out}^2))\big|\\
  &\phantom{\leq}
  +\big|C_\mathrm{sum}(\mathbf{C}_\mathrm{in},(C_\mathrm{out}^1,\tilde{C}_\mathrm{out}^2))
  -C_\mathrm{sum}(\mathbf{C}_\mathrm{in},(C_\mathrm{out}^1,0))\big|,
\end{align*}
where (a) follows from the triangle inequality, and (b) follows from Lemma
\ref{lem:concaveFunctions}. If we now let 
$\mathbf{\tilde{C}}_\mathrm{out}\rightarrow (C_\mathrm{out}^1,0)$, 
Corollary \ref{cor:continuityAtZero} implies
\begin{align*}
  \MoveEqLeft
  \lim_{\mathbf{\tilde{C}}_\mathrm{out}\rightarrow (C_\mathrm{out}^1,0)}
  \big|C_\mathrm{sum}(\mathbf{C}_\mathrm{in},(\tilde{C}_\mathrm{out}^1,\tilde{C}_\mathrm{out}^2))
  -C_\mathrm{sum}(\mathbf{C}_\mathrm{in},(C_\mathrm{out}^1,0))\big|\\
  &\leq
  \lim_{\tilde{C}_\mathrm{out}^2\rightarrow 0}
  \big|C_\mathrm{sum}(\mathbf{C}_\mathrm{in},(C_\mathrm{out}^1,\tilde{C}_\mathrm{out}^2))
  -C_\mathrm{sum}(\mathbf{C}_\mathrm{in},(C_\mathrm{out}^1,0))\big|,
\end{align*}
from which our result follows. An analogous proof applies
in the case where $C_\mathrm{out}^1=0$, but 
$C_\mathrm{out}^2>0$.

\section{Summary}
Consider a network consisting of a discrete MAC
and a CF that has full knowledge of the messages.
In this work, we show that the average-error
sum-capacity of such a network is always a continuous
function of the CF output link capacities; this 
is in contrast to our previous results on maximal-error sum-capacity
\cite{reliability}.
Our proof method relies on finding lower and upper bounds
on the average-error sum-capacity and then using a modified
version of a technique developed by Dueck \cite{DueckSC} 
to demonstrate continuity.

\bibliographystyle{IEEEtran}
\bibliography{ref}{}

\end{document}